\documentclass[preprintnumbers,two column, showpacs,amsmath]{revtex4}
\usepackage{graphicx}
\usepackage{dcolumn}
\usepackage{bm}

\begin{document}

\title{Exact solvability  of the quantum Rabi models within Bogoliubov operators}
\author{Qing-Hu Chen$^{1,2}$, Chen Wang$^{1}$, Shu He$^{1,3}$, Tao Liu$^{3}$, and Ke-Lin Wang$^{4}$}

\address{
$^{1}$  Department of Physics, Zhejiang University, Hangzhou 310027, P. R. China \\
$^{2}$ Center for Statistical and Theoretical Condensed Matter
Physics, Zhejiang Normal University, Jinhua 321004, P. R.
China \\
$ ^{3}$ School of Science,  Southwest University of  Science and
Technology, Mianyang 621010, P.  R.  China\\
$^{4}$ Department of Modern Physics, University of  Science and
Technology of China,  Hefei 230026, P.  R.  China
 }\date{\today}

\begin{abstract}
The quantum Rabi model can be solved exactly by the Bargmann
transformation from real coordinate to complex variable recently
[Phys. Rev. Lett. \textbf{ 107}, 100401 (2011)]. By the extended
coherent states, we recover this solution in an alternative simpler
and perhaps more physical way without uses of any extra conditions,
like  Bargmann conditions. In the same framework, the two-photon
Rabi model are solved exactly by extended squeeze states.
Transcendental functions have been derived with the similar form as
those in one-photon model. Both extended coherent states and squeeze
states are essentially Fock states in the space of the corresponding
Bogoliubov operators. The present approach could be easily extended
to study the exact solvability or integrability of various
spin-boson systems with multi-level, even multi-mode.
\end{abstract}

\pacs{03.65.Ge, 42.50.Pq  42.50.Lc} \maketitle

\section{Introduction}

Matter-light interaction is fundamental and ubiquitous in modern physics
ranging from quantum optics, quantum information science to condensed matter
physics. The simplest paradigm is a two-level atom (qubit) coupled to the
electromagnetic mode of a cavity (oscillator), which is called Rabi model
\cite{Rabi}. In the strong coupling regime where the coupling strength $%
g/\omega $ ($\omega $ is the cavity frequency) between the atom and the
cavity mode exceeds the loss rates, the atom and the cavity can repeatedly
exchange excitations before coherence is lost. The Rabi oscillations can be
observed in this strong coupling atom-cavity system, which is usually called
as cavity quantum electrodynamics (QED) \cite{CQED}. Typically, the coupling
strength in cavity QED reaches $g/\omega \sim 10^{-6}$. It can be described
by the Jaynes-Cummings (JC) model~ \cite{JC} where the rotating-wave
approximation (RWA) is made and analytically closed-form exact solutions are
available.

Recently, for superconducting qubits, a one-dimensional (1D) transmission
line resonator~\cite{Wallraff} or a LC circuit~\cite
{Chiorescu,Wang,Deppe,Fink} can play a role of the cavity, which is known
today as circuit QED. More recently, LC resonator inductively coupled to a
superconducting qubit~\cite{Niemczyk,exp,Mooij} has been realized
experimentally. The qubit-resonator coupling has been strengthened to ten
percentage. In this ultrastrong coupling regime of circuit QED, the evidence
for the breakdown of the RWA has been provided by the transmission spectra~%
\cite{Niemczyk}. The remarkable Bloch-Siegert shift associated with the
counter-rotating terms also demonstrates the failure of the RWA~\cite{exp}.
So the quantum Rabi model (QRM) has been revisited by many authors.

In the representation of bosonic creation and annihilation operators in the
Bargmann space \cite{Bargmann} of analytical functions in a complex
variable, Braak \cite{Braak} recently derived a transcendental function in
the QRM, which is defined through power series in the coupling strength with
coefficients related recursively. Zeros of transcendental functions can give
exact eigenvalues. By a proposed criterion for quantum integrability, Braak
further shows that the QRM is integrable due to the parity symmetry.
However, the derivations seems to be outlined in a mathematical way. It was
also suggested \cite{Solano} that an intense dialogue between mathematics
and physics would be needed. In other words, it is useful to shed some
physical insights into Braak's mainly mathematical solutions.

In this paper, without the use of any extra conditions, like analyticity of
the eigenfunction in Bargmann representation, we alternatively re-derive the
same transcendental functions as in Ref. \cite{Braak} quantum mechanically
within the extended coherent stats\cite{chenqh,Qinghu}. Both zero bias and
biased QRM can be treated simultaneously. Our method is more intuitionistic,
and may be more easily understandable. The key procedures can be described
with a simple tutorial, and therefore it is straightforward to extend to
study the exact solvability or integrability of various spin-boson systems.
The extension to the two-photon QRM\cite{Albert,Trav} is performed in this
paper as the first example.

\section{The QRM within Bogoliubov operators}

\subsection{Re-derivation of Braak's solution}

The Hamiltonian of a generalized QRM can be describe as follows
\begin{equation}
H=-\frac 12\left( \varepsilon \sigma _z+\Delta \sigma _x\right) +a^{\dagger
}a+g(a^{\dagger }+a)\sigma _z,  \label{Hamiltonian}
\end{equation}
where $\varepsilon $ and $\Delta $ are qubit static bias and tunneling
matrix element, $a^{\dagger }$, $a$ are the photon creation and annihilation
operators of the single-mode cavity with frequency $\omega $, $g$ is the
qubit-cavity coupling constant, and $\sigma _k(k=x,y,z)$ $\ $ are the Pauli
matrices. To facilitate the study, we write the Hamiltonian in the matrix
form in units of $\hbar =\omega =1$
\begin{equation}
H=\left(
\begin{array}{ll}
a^{\dagger }a+g\left( a^{\dagger }+a\right) -\frac \varepsilon 2 &
~~~~~~~~-\frac \Delta 2 \\
~~~~~~~~-\frac \Delta 2 & a^{\dagger }a-g\left( a^{\dagger }+a\right) +\frac
\varepsilon 2
\end{array}
\right) .
\end{equation}
To remove the linear terms in $a^{\dagger }(a)$ a operators, we perform the
following two Bogoliubov transformations
\begin{equation}
A=a+g,B=a-g.
\end{equation}
In Bogoliubov operators $A$($B$), the matrix element $H_{11}$ ($H_{22}$) can
be reduced to the free particle number operators $A^{\dagger }A$ ($%
B^{\dagger }B$) plus a constant, which is very helpful for the further study.

Different from the previous ansatz that the Hamiltonian is expressed in the
two operators $A$ and $B$ at the same time\cite{chenqh}, we here use the
single operator one by one. First, in terms of operator $A$, the Hamiltonian
can be written as
\begin{equation}
H=\left(
\begin{array}{ll}
A^{\dagger }A-\alpha  & ~~~~~~~~-\frac \Delta 2 \\
~-\frac \Delta 2 & A^{\dagger }A-2g\left( A^{\dagger }+A\right) +\beta
\end{array}
\right) ,
\end{equation}
where
\[
\alpha =g^2+\frac \varepsilon 2,\;\beta =3g^2+\frac \varepsilon 2.
\]
The wavefunction is then proposed as
\begin{equation}
\left| {}\right\rangle =\left( \
\begin{array}{l}
\sum_{n=0}^\infty \sqrt{n!}e_n\left| n\right\rangle _A \\
\sum_{n=0}^\infty \sqrt{n!}f_n\left| n\right\rangle _A
\end{array}
\right) ,  \label{wave1}
\end{equation}
where $e_n\;$and $f_n$ are the expansion coefficients, $\left|
n\right\rangle _A\;$is called extended coherent state with the following
properties
\begin{eqnarray}
\left| n\right\rangle _A &=&\frac{\left( A^{\dagger }\right) ^n}{\sqrt{n!}}%
\left| 0\right\rangle _A=\frac{\left( a^{\dagger }+g\right) ^n}{\sqrt{n!}}%
\left| 0\right\rangle _A,  \label{ex1} \\
\left| 0\right\rangle _A &=&e^{-\frac 12g^2-ga^{\dagger }}\left|
0\right\rangle _a.  \label{ex2}
\end{eqnarray}
The Schr$\stackrel{..}{o}$dinger equation gives
\begin{eqnarray*}
&&\sum_{n=0}^\infty \left( n-\alpha \right) \sqrt{n!}e_n\left|
n\right\rangle _A-\frac \Delta 2\sum_{n=0}^\infty \sqrt{n!}f_n\left|
n\right\rangle _A \\
&=&E\sum_{n=0}^\infty \sqrt{n!}e_n\left| n\right\rangle _A
\end{eqnarray*}
\begin{eqnarray*}
&&-\frac \Delta 2\sum_{n=0}^\infty \sqrt{n!}e_n\left| n\right\rangle
_A+\sum_{n=0}^\infty \left( n+\beta \right) \sqrt{n!}f_n\left|
n\right\rangle _A \\
&&-2g\sum_{n=0}^\infty \left( \sqrt{n}f_n\sqrt{n!}\left| n-1\right\rangle _A+%
\sqrt{n+1}\sqrt{n!}f_n\left| n+1\right\rangle _A\right)  \\
&=&E\sum_{n=0}^\infty \sqrt{n!}f_n\left| n\right\rangle _A
\end{eqnarray*}
Left multiplying $_A\left\langle m\right| $ gives
\begin{equation}
\left( m-\alpha -E\right) e_m=\frac \Delta 2f_m,
\end{equation}
\begin{equation}
\left( m+\beta -E\right) f_m-2g\left( m+1\right) f_{m+1}-2gf_{m-1}=\frac
\Delta 2e_m.
\end{equation}
So the two coefficients $e_n\;$and $f_n\;$ with the same index $n$ are
related with
\begin{equation}
e_m=\frac \Delta {2\left( m-\alpha -E\right) }f_m,  \label{cor1}
\end{equation}
and the coefficient $f_n$ can be defined recursively,
\begin{eqnarray}
mf_m &=&\Omega (m-1)f_{m-1}-f_{m-2}, \\
\Omega (m) &=&\frac 1{2g}\left( \left( m+\beta -E\right) -\frac{\Delta ^2}{%
4\left( m-\alpha -E\right) }\right) ,
\end{eqnarray}
with\ $f_0=1\;$and $f_1=\Omega (0)$

Similarly, by the second operator $B$, we can have the Hamiltonian as
\begin{equation}
H=\left(
\begin{array}{ll}
B^{\dagger }B+2g\left( B^{\dagger }+B\right) +\beta ^{\prime } & ~~-\frac
\Delta 2 \\
~~~~~~~~~~~~-\frac \Delta 2 & B^{\dagger }B-\alpha ^{\prime }
\end{array}
\right) ,
\end{equation}
where
\[
\alpha ^{\prime }=g^2-\frac \varepsilon 2,\;\;\beta ^{\prime }=3g^2-\frac
\varepsilon 2.
\]
The wavefunction can be also suggested in terms of $B$ as
\begin{equation}
\left| {}\right\rangle =\left( \
\begin{array}{l}
\sum_{n=0}^\infty \left( -1\right) ^n\sqrt{n!}f_n^{\prime }\left|
n\right\rangle _B \\
\sum_{n=0}^\infty \left( -1\right) ^n\sqrt{n!}e_n^{\prime }\left|
n\right\rangle _B
\end{array}
\right) .  \label{wave2}
\end{equation}
Proceed as before, the two coefficients $f_n^{\prime }\;$and $e_n^{\prime }$
satisfy
\[
-2g\left( f_{m-1}^{\prime }+\left( m+1\right) f_{m+1}^{\prime }\right)
+\left( m+\beta ^{\prime }-E\right) f_m^{\prime }-\frac \Delta 2e_m^{\prime
}=0,
\]
\[
-\frac \Delta 2f_m^{\prime }+\left( m-\alpha ^{\prime }\right) e_m^{\prime
}=Ee_m^{\prime },
\]
then we have
\begin{equation}
e_m^{\prime }=\frac{\frac \Delta 2}{m-\alpha ^{\prime }-E}f_m^{\prime },
\label{cor2}
\end{equation}
\begin{eqnarray}
mf_m^{\prime } &=&\Omega ^{\prime }(m-1)f_{m-1}^{\prime }-f_{m-2}^{\prime },
\\
\Omega ^{\prime }(m) &=&\frac 1{2g}\left[ \left( m+\beta ^{\prime }-E\right)
-\frac{\Delta ^2}{4\left( m-\alpha ^{\prime }-E\right) }\right] ,
\end{eqnarray}
with $f_0^{\prime }=1\;$and$\;f_1^{\prime }=\Omega ^{\prime }(0).$

If both wavefunctions (\ref{wave1}) and (\ref{wave2}) are the true
eigenfunction for the same eigenvalue $E$, they should be in principle only
different by a complex constant $r$ if this eigenvalue is not degenerate
\begin{eqnarray}
\sum_{n=0}^\infty \sqrt{n!}e_n\left| n\right\rangle _A &=&r\sum_{n=0}^\infty
(-1)^n\sqrt{n!}f_n^{\prime }\left| n\right\rangle _B, \\
\sum_{n=0}^\infty \sqrt{n!}f_n\left| n\right\rangle _A &=&r\sum_{n=0}^\infty
(-1)^n\sqrt{n!}e_n^{\prime }\left| n\right\rangle _B.
\end{eqnarray}
Left multiplying the original vacuum state ${\langle }0|$ to the both side
of the above equations yields
\begin{eqnarray}
\sum_{n=0}^\infty \sqrt{n!}e_n{\langle }0|\left| n\right\rangle _A
&=&r\sum_{n=0}^\infty (-1)^n\sqrt{n!}f_n^{\prime }{\langle }0|\left|
n\right\rangle _B, \\
\sum_{n=0}^\infty \sqrt{n!}f_n{\langle }0|\left| n\right\rangle _A
&=&r\sum_{n=0}^\infty (-1)^n\sqrt{n!}e_n^{\prime }{\langle }0|\left|
n\right\rangle _B,
\end{eqnarray}
where
\begin{equation}
\sqrt{n!}{\langle }0|n{\rangle }_A=(-1)^n\sqrt{n!}{\langle }0|n{\rangle }%
_B=e^{-g^2/2}g^n.
\end{equation}
To eliminate the ratio constant $r$, we have the following relation
\[
\sum_{n=0}^\infty e_ng^n\sum_{n=0}^\infty e_n^{\prime }g^n=\sum_{n=0}^\infty
f_ng^n\sum_{n=0}^\infty f_n^{\prime }g^n,
\]
with the help of Eqs. (\ref{cor1}) and (\ref{cor2}), we arrive at
\begin{eqnarray}
&&\sum_{n=0}^\infty \frac{\Delta /2}{n-\alpha -E}f_ng^n\sum_{n=0}^\infty
\frac{\Delta /2}{n-\alpha ^{\prime }-E}f_n^{\prime }g^n  \nonumber \\
&=&\sum_{n=0}^\infty f_ng^n\sum_{n=0}^\infty f_n^{\prime }g^n.
\end{eqnarray}
If set $f_n{=}K_n^{-}$, $f_n^{^{\prime }}{=}K_n^{+}$ and $E=x-g^2$, we can
recover Braak's exact solution\cite{Braak}
\begin{equation}
G_\epsilon (x)=(\frac \Delta 2)^2\overline{R}^{+}(x)\overline{R}%
^{-}(x)-R^{+}(x)R^{-}(x)=0,  \label{bias}
\end{equation}
where
\begin{eqnarray*}
R^{\pm }(x) &=&\sum_{n=0}^\infty K_n^{\pm }(x)\;g^n, \\
\overline{R}^{\pm }(x) &=&\sum_{n=0}^\infty \frac{K_n^{\pm }(x)}{x-n\mp
\frac \varepsilon 2}\;g^n.
\end{eqnarray*}
If $\varepsilon =0$, the above equation can be reduced to the following
zero-bias case obviously\cite{Braak}
\begin{equation}
G_0^{\pm }(x)=\sum_{n=0}^\infty f_n(x)\left( 1\mp \frac{\Delta /2}{x-n}%
\;\right) g^n=0.  \label{unbias}
\end{equation}
Therefore Braak's G-functions are completely re-derived in a very
intuitionistic and concise way.

The G-functions can be written \cite{Braak'commu} in terms of
so-called Heun functions \cite{book}. However, the zeros of this
Heun functions can not be given analytically, the numerical
technique in the search for the zeros is still needed, so the
cut-off for the summation should be unavoidable in the practical
evaluation.

\subsection{Comparisons and discussions}

In the above derivation, the key point is proportionality of the two
wavefunctions (\ref{wave1} ) and (\ref{wave2}) with the same
eigenvalue. Both Hilbert spaces in the two Bogoliubov operators are
complete, if truncation is not done, the proportionality is
justified naturally for non-degenerate states. On the other hand,
the degenerate eigenstates are excluded in principle here. It
naturally follows that the Juddian solutions \cite{Judd,Koc}, which
eigenvalue is doubly degenerate, are exceptional to Braak's
solution. Interestingly, we do not need the extra condition of
analyticity of the eigenfunction in Bargmann representation. In
addition, the validity of present approach is independent of the
parity symmetry. The parity symmetry would be contained
self-consistently in the final G-functions if the system really has,
e.g. $\varepsilon =0$.

Based on two Bogoliubov operator $A$ and $B$, three present authors and one
collaborator have used the following wavefunction to the Hamiltonian(\ref
{Hamiltonian}) to analyze the spectrum in the qubit-oscillator systems [c.f.
Eq. (6) in Ref. \cite{Qinghu}]
\begin{equation}
\left| {}\right\rangle =\left(
\begin{array}{l}
\sum_{n=0}^Nc_n\left| n\right\rangle _A \\
\sum_{n=0}^Nd_n\left| n\right\rangle _B
\end{array}
\right) ,  \label{wavefunction_bias}
\end{equation}
where $N$ is the truanted number. The numerical exact diagonalization (ED)
in the space of the two Bogoliubov operators have given the spectrum
exactly. The coefficients $c_n\;$and $d_n\ $can be obtained also.

It is interesting to link coefficients in wavefunction (\ref
{wavefunction_bias}) and those in wavefunctions (\ref{wave1}) and (\ref
{wave2}) as
\begin{eqnarray*}
\;\;c_n &=&\sqrt{n!}e_n, \\
d_n &=&r\left( -1\right) ^n\sqrt{n!}e_n^{\prime },
\end{eqnarray*}
although the former ones are obtained from ED and the later ones by the
zeros of the G functions. It can be also confirmed numerically. For
practical interest, there are perhaps no essential differences between our
work and Braak's solution\cite{Braak}, except that the avenues to obtain the
basically same coefficients are different. The latter is described in a
mathematical way and is of more conceptual interest.

In the mathematical sense, we can not rule out the possibility that the ED
give good results for small $N$ and get worse for higher $N$ before the
practical evaluation, although we know empirically that it is generally not
that case for large $N$. For the low order perturbation theory, it happens
that the third-order perturbation theory would give worse results than the
second-order one in some parameter regime for instance, it may be not that
case in very high order perturbation theory. In the calculation, we really
find that relative difference between the exact ones, which are easily
obtained to any desired accuracy, and those for the cut off $N$ decreases
monotonically with increasing $N$, and the convergence can be arrived at
easily. One may know that the Heun series converges before numerical
calculations, although the cut-off can not be avoided in the real
calculation. We indeed did not unfold the mathematics behind this formalism
previously and focused on the analysis of the experimental spectrum data at
that time\cite{exp}.

\begin{figure}[tbp]
\includegraphics[width=8cm]{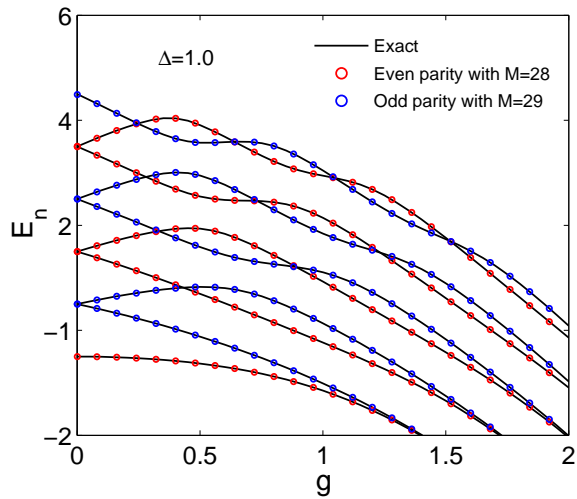}
\caption{(Color online) Spectrum of the one-photon quantum Rabi
model as zeros of functions $F(\alpha)$ in Eq. (\ref{ours}) at
resonance. The numerical solutions are also collected with solid
lines. } \label{Spectrum_0}
\end{figure}

Braak's G-functions exhibit a very compact form in power series, which
motivate us to reshape our previous work. By tunable extended bosonic
coherent states, the QRM can be mapped to a polynomial equation with a
single variable\cite{QingHu1}. We can also write this polynomial in power
series as the following more concise form for large truanted number $M$
\begin{equation}
F(\alpha )=\sum_{j=0}^M\frac{(2\alpha )^j}{j!}c_{_{M-j}}=0,  \label{ours}
\end{equation}
where $\alpha $ is the key tunable variable we seek, and coefficients are
also related recursively with the following scheme
\begin{eqnarray}
(m+1)gc_{m+1} &=&-(m\pm \frac \Delta 2)c_m-\left( \alpha +g\right) c_{m-1}
\nonumber \\
&&\pm (-1)^m\frac \Delta 2\sum_{j=0}^m\frac{(2\alpha )^j}{j!}c_{m-j}],
\end{eqnarray}
initiated from $c_0=1.0$ and $c_1=0$, because the coefficients with the two
highest indices $M$ and $M-1$ are negligible small due to the required
convergence and can thus be omitted. The zeros of the above function $%
F(\alpha )$ can also give the exact eigenvalue through
\begin{equation}
E^{\pm }=\alpha g\mp \frac \Delta 2.  \label{zero}
\end{equation}
where $\pm $ denotes the parity. The results are shown in Fig. \ref
{Spectrum_0}. In Eq. (20) of the end of Ref. \cite{QingHu1}, we have
demonstrated that the wavefunction is equivalent to the expansion in the
Fock space of displaced operators with tunable displacement.

It is very interesting to note that zeros of the both functions defined
through different power series in Eqs. ( \ref{unbias}) and (\ref{ours}) can
give the exact eigenvalues. In our practical evaluation, it is not more
difficult to find the zeros for the function in Eq. (\ref{ours}) than those
in Eq. ( \ref{unbias}), because the poles at $x=n$ emerging in the latter
are not present in the former. The key difference between Eqs. ( \ref{unbias}%
) and (\ref{ours}) is that the former is well-defined without restriction
and the latter is well-defined with build-in truncation.

It is implied in the viewpoint\cite{Solano} that the QRM might have been
solved exactly with an analytical closed-form solution in Ref. \cite{Braak}.
Nevertheless, whether Braak's exact solution could be called closed-form is
subtle and therefore still controversial in our opinion. The so-called Heun
functions can be basically called closed-form because they are well defined,
although much more complicated than e.g. the hypergeometric functions. But
the eigenvalues are given by the zeros of the Heun functions, which can not
be obtained without truncation in the power series. As shown in
wavefunctions (\ref{wave1} ) and (\ref{wave2}), the expansion can not be
closed naturally like in the JC model. It is generally accepted that the QRM
has no trivial closed-form solution like that in the JC model due to the
counter-rotating terms. The QRM can only have closed-form solutions with a
vanishing qubit tunneling $\Delta =0$\cite{QingHu1,Casanova}. Perhaps the
question of "closed-form" solutions is academic and not of real value.
Braak's solution is interesting for integrability of the QRM.

\subsection{Tutorial for Bogoliubov operators approach}

The present approach within the Bogoliubov operators can be generally
described as follows, which is helpful for the further applications. The
main task is to find the corresponding Bogoliubov operators. Then, one can
expand the wavefunctions in terms of each Bogoliubov operator respectively.
Eliminating the ratio constant of these wavefunctions will give the
transcendental functions, which would be defined through power series in
model parameter dependent quantities with coefficients related recursively.
Finally, zeros of these transcendental functions would give the eigenvalues
exactly, where numerical solutions to the one-variable (or finite variables
in other multi-level systems for example) nonlinear equation must be
required. Although the power series are defined through an infinite
summation formally, in the practical calculation, it should be truncated to
a finite summation. Fortunately, the obtained transcendental function $G(x)$
can be written in terms of so-called Heun functions, by which we know the
convergence before the numerical solutions. The unavoidable "cut-off" in the
summation of the G-functions in the practical calculations means that some
states in the Hilbert space is not considered, according to the
wavefunctions (\ref{wave1}) and (\ref{wave2}), even they contribute
negligible small, they still belong to the Hilbert space of Bogoliubov
operators.

The applications to the two-photon QRM is performed in the next
section as a preliminary example.

\section{Extensions to the two-photon QRM}

The two-photon QRM is an phenomenological model describing a three-level
system interacting with two photons. When the intermediate transition
frequencies are strongly detuned from cavity frequency, after adiabatically
eliminating the intermediate levels, one arrives at the effective
Hamiltonian. This two-photon QRM has also been studied for a long time with
the RWA\cite{rwa} and beyond the RWA\cite{Toor,Albert,Trav}. It may describe
the two-photon processes occurring in Rubidium atoms\cite{Brune} and quantum
dots\cite{Valle}.

The Hamiltonian of two-photon QRM takes the following matrix form
\begin{equation}
H=\left(
\begin{array}{ll}
a^{\dagger }a+g\left[ \left( a^{\dagger }\right) ^2+a^2\right] &
~~~~~~~~-\frac \Delta 2 \\
~~~~~~~~-\frac \Delta 2 & a^{\dagger }a-g\left[ \left( a^{\dagger }\right)
^2+a^2\right]
\end{array}
\right).
\end{equation}

First, we perform Bogoliubov transformation
\begin{equation}
b=ua+va^{\dagger },b^{\dagger }=ua^{\dagger }+va,
\end{equation}
to generate a new bosonic operator. Compared to the Hamiltonian, if set
\begin{equation}
u=\sqrt{\frac{\beta +1}2},\;v=\sqrt{\frac{\beta -1}2},
\end{equation}
with $\beta =\frac 1{\sqrt{1-4g^2}}$, we have a simple quadratic form of one
diagonal Hamiltonian matrix element
\[
H_{11}=a^{\dagger }a+g\left[ \left( a^{\dagger }\right) ^2+a^2\right] =\frac{%
b^{\dagger }b-v^2}{u^2+v^2}.
\]
Similarly, we can introduce another operator
\begin{equation}
c=ua-va^{\dagger },c^{\dagger }=ua^{\dagger }-va,
\end{equation}
which yields a simple quadratic form of the other diagonal Hamiltonian
matrix element
\[
H_{22}=a^{\dagger }a-g\left[ \left( a^{\dagger }\right) ^2+a^2\right] =\frac{%
c^{\dagger }c-v^2}{u^2+v^2}.
\]

In terms of the Bogoliubov operator $b$, the Hamiltonian can be written as
\begin{equation}
H=\left(
\begin{array}{ll}
\frac{b^{\dagger }b-v^2}{u^2+v^2}-v^2 & ~-\frac \Delta 2 \\
~~-\frac \Delta 2 & \;\;\;H_{22}^{\prime }
\end{array}
\right),
\end{equation}
with
\begin{eqnarray*}
&&H_{22}^{\prime }=\left( u^2+v^2+4guv\right) b^{\dagger }b \\
&&-\left[ uv+g\left( u^2+v^2\right) \right] \left[ \left( b^{\dagger
}\right) ^2+b^2\right] +2guv+v^2.
\end{eqnarray*}
The wavefucntions is suggested as
\begin{equation}
\left| {}\right\rangle =\left( \
\begin{array}{l}
\sum_{n=0}\sqrt{n!}e_n\left| n\right\rangle _b \\
\sum_{n=0}\sqrt{n!}f_n\left| n\right\rangle _b
\end{array}
\right),
\end{equation}
where
\begin{eqnarray}
\left| n\right\rangle _b &=&\frac{\left( b^{\dagger }\right)
^n}{\sqrt{n!}} \left| 0\right\rangle _b=\frac{\left( ua^{\dagger
}+va\right) ^n}{\sqrt{n!}}
\left| 0\right\rangle _b, \\
b\left| 0\right\rangle _b &=&0,
\end{eqnarray}
$\left| 0\right\rangle _b$ is simply the single-mode squeezed vacuum
state, $ \left| n\right\rangle _b\;$is thus called as extended
squeezed state. The Schr$\stackrel{..}{o}$dinger equation gives
\begin{eqnarray*}
&&\sum_{n=0}\frac{b^{\dagger }b-v^2}{u^2+v^2}\sqrt{n!}e_n\left|
n\right\rangle _b-\frac \Delta 2\sum_{n=0}\sqrt{n!}f_n\left| n\right\rangle
_b \\
&=&E\sum_{n=0}\sqrt{n!}e_n\left| n\right\rangle _b,
\end{eqnarray*}
\begin{eqnarray*}
&&\left( u^2+v^2+4guv\right) b^{\dagger }b\sum_{n=0}\sqrt{n!}f_n\left|
n\right\rangle _b \\
&&-\left[ uv+g\left( u^2+v^2\right) \right] \left[ \left( b^{\dagger
}\right) ^2+b^2\right] \sum_{n=0}\sqrt{n!}f_n\left| n\right\rangle _b \\
&&+\left( 2guv+v^2\right) \sum_{n=0}\sqrt{n!}f_n\left| n\right\rangle
_b-\frac \Delta 2\sum_{n=0}\sqrt{n!}e_n\left| n\right\rangle _b \\
&=&E\sum_{n=0}\sqrt{n!}f_n\left| n\right\rangle _b.
\end{eqnarray*}
Left multiplying $_b\left\langle m\right| $ gives
\[
\left( \frac{m-v^2}{u^2+v^2}-E\right) e_m-\frac \Delta 2f_m=0,
\]
\begin{eqnarray*}
&&-\left[ uv+g\left( u^2+v^2\right) \right] \left[ f_{m-2}+\left( m+2\right)
\left( m+1\right) f_{m+2}\right]  \\
&&+\left[ \left( u^2+v^2\right) m+v^2+2guv\left( 2m+1\right) -E\right] f_m \\
&&-\frac \Delta 2e_m=0.
\end{eqnarray*}
Then we have build one-by-one relation for coefficient $e_m$ and $f_m$
\begin{equation}
e_m=\frac \Delta {2\left[ \frac{m-v^2}{u^2+v^2}-E\right] }f_m,
\label{coeef1}
\end{equation}
which will considerably simplify the problem. The reclusive relation is then
obtained as
\begin{equation}
\left( m+2\right) \left( m+1\right) f_{m+2}=-f_{m-2}+\frac{\Omega (m)}{%
uv+g\left( u^2+v^2\right) }f_m  \label{recursive_2},
\end{equation}
where
\begin{eqnarray}
\Omega (m) &=&\left( u^2+v^2\right) m+v^2+2guv\left( 2m+1\right),
\nonumber
\\
&&-E-\frac{\Delta ^2}{4\left( \frac{m-v^2}{u^2+v^2}-E\right) }.
\end{eqnarray}

The Hamiltonian can also be expressed in terms of the other Bogoliubov
operator $c$
\begin{equation}
H=\left(
\begin{array}{ll}
\;H_{11}^{\prime } & -\frac \Delta 2 \\
~-\frac \Delta 2\; & \frac{c^{\dagger }c-v^2}{u^2+v^2}
\end{array}
\right),
\end{equation}
with
\begin{eqnarray*}
H_{11}^{\prime } &=&\left( v^2+u^2+4guv\right) c^{\dagger }c \\
&&+\left[ uv+g\left( v^2+u^2\right) \right] \left[ \left( c^{\dagger
}\right) ^2+c^2\right] +2guv+v^2.
\end{eqnarray*}
The wavefunction then can be expanded in the Fock space of the $c$ operator
as the following form
\begin{equation}
\left| {}\right\rangle =\left( \
\begin{array}{l}
\sum_{n=0}i^l\sqrt{n!}f_n^{\prime }\left| n\right\rangle _c \\
\sum_{n=0}i^l\sqrt{n!}e_n^{\prime }\left| n\right\rangle _c
\end{array}
\right),
\end{equation}
where $l=n$ for $n=2k$ and $l=n+1$ for $n=2k+1$. Therefore only two values
of $i^l=\pm 1$ are possible.

Similarly, the Schr$\stackrel{..}{o}$dinger equation gives the following
relations
\[
-\frac \Delta 2f_m^{\prime }+\frac{m-v^2}{u^2+v^2}e_m^{\prime
}=Ee_m^{\prime },
\]
\begin{eqnarray*}
&&-\left[ g\left( v^2+u^2\right) +uv\right] \left[ (m+2)\left(
m+1\right)
f_{m+2}^{\prime }+f_{m-2}^{\prime }\right] \\
&&+\left[ v^2+\left( u^2+v^2\right) m+2guv\left( 2m+1\right)
-E\right] f_m^{\prime }-\frac \Delta 2e_m^{\prime }=0.
\end{eqnarray*}
The coefficient $e_m$ and $f_m$ are related by
\begin{equation}
e_m^{\prime }=\frac \Delta {2\left[ \frac{m-v^2}{u^2+v^2}-E\right]
}f_m^{\prime },  \label{coeef2}
\end{equation}
and the reclusive relation is
\begin{equation}
(m+2)\left( m+1\right) f_{m+2}^{\prime }=-f_{m-2}^{\prime
}+\frac{\Omega ^{\prime }(m)}{uv+g\left( u^2+v^2\right) }f_m^{\prime
},
\end{equation}
with
\begin{eqnarray}
\Omega ^{\prime }(m) &=&v^2+\left( u^2+v^2\right) m+2guv\left( 2m+1\right)
\nonumber \\
&&-E-\frac{\Delta ^2}{4\left( \frac{m-v^2}{u^2+v^2}-E\right) }.
\end{eqnarray}
Note that the two sets of coefficients in the two wavefucntions has the same
form.

\begin{figure}[tbp]
\includegraphics[width=8cm]{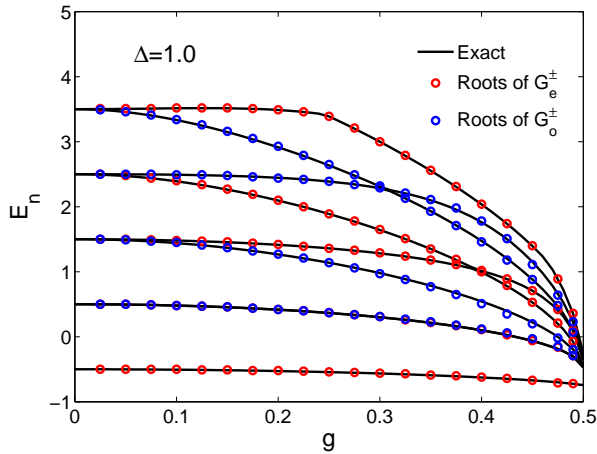}
\caption{(Color online) Spectrum of the two-photon quantum Rabi
model as zeros of G-functions in Eq. (\ref{cerntral_2p}). The
numerical solutions are also collected with solid lines. }
\label{Spectrum}
\end{figure}

Similarly, the two wavefunctions with the same eigenvalue should be
in principle proportional with each other for the non-degenerate
state
\begin{equation}
\left( \
\begin{array}{l}
\sum_{n=0}\sqrt{n!}e_n\left| n\right\rangle _b \\
\sum_{n=0}\sqrt{n!}f_n\left| n\right\rangle _b
\end{array}
\right) =r\left( \
\begin{array}{l}
\sum_{n=0}i^l\sqrt{n!}f_n^{\prime }\left| n\right\rangle _c \\
\sum_{n=0}i^l\sqrt{n!}e_n^{\prime }\left| n\right\rangle _c
\end{array}
\right).
\end{equation}
Left multiplying ${\langle }0|$ to the both equations gives
\begin{eqnarray*}
\sum_{n=0}\sqrt{n!}e_n{\langle }0\left| n\right\rangle _b &=&r\sum_{n=0}i^l%
\sqrt{n!}f_n^{\prime }{\langle }0\left| n\right\rangle _c, \\
\sum_{n=0}\sqrt{n!}f_n{\langle }0\left| n\right\rangle _b &=&r\sum_{n=0}i^l%
\sqrt{n!}e_n^{\prime }{\langle }0\left| n\right\rangle _c.
\end{eqnarray*}
We always have
\begin{equation}
i^l\sqrt{n!}{\langle }0\left| n\right\rangle _c=\sqrt{n!}{\langle
}0\left| n\right\rangle _b=L_n^{e,o},
\end{equation}
where
\begin{eqnarray}
L_{n=2k}^e &=&\frac{\left( 2k\right) !(uv)^k}{2^k}\sum_{j=0}^k\frac{(-\frac{%
v^2}{u^2})^j}{j!(k-j)!}, \\
L_{n=2k+1}^o &=&\frac{\left( 2k+1\right) !v(uv)^k}{2^k}\sum_{j=0}^k\frac{%
2^{2j}j!(-\frac{v^2}{u^2})^j}{\left( 2j+1\right) !(k-j)!},
\end{eqnarray}
for even and odd particle numbers in the Bogoliubov operators $b$ and $c$
respectively. Then we have
\[
\sum_ne_nL_n^{e,o}=r\sum_nf_n^{\prime
}L_n^{e,o};\;\sum_nf_nL_n^{e,o}=r\sum_ne_n^{\prime }L_n^{e,o}.
\]
Now the summation $\sum_n$ is separated into two series with even and odd
number $n$. To eliminate the constant $r$, we have
\begin{eqnarray}
&&\sum_n\frac \Delta {2\left( \frac{n-v^2}{u^2+v^2}-E\right)
}f_nL_n^{e,o}\sum_n\frac \Delta {2\left( \frac{n-v^2}{u^2+v^2}-E\right)
}f_n^{\prime }L_n^{e,o}  \nonumber \\
&=&\sum_nf_n^{\prime }L_n^{e,o}\sum_nf_nL_n^{e,o}, \label{protion}
\end{eqnarray}
with the use of Eqs. (\ref{coeef1}) and (\ref{coeef2}). Set $f_n=f_n^{\prime
}$ and $-x=-v^2-E\left( u^2+v^2\right) $, we finally have
\begin{equation}
G_{e,o}^{\pm }=\sum_nf_n\;\left[ 1\pm \frac{\Delta \left( u^2+v^2\right) }{%
2\left( n-x\right) }\;\right] L_n^{e,o}=0,  \label{cerntral_2p}
\end{equation}
where the coefficient $f_n$ is initiated from $f_0=1$ ($f_1=1$) for the case
of the even (odd) $n$ in the recurrence scheme Eq. (\ref{recursive_2}), and $%
\pm $ denotes the parity. Thus, G-functions for the two-photon QRM have been
obtained. The zeros of the G-functions give the exact eigenvalues, as shown
in Fig. \ref{Spectrum}. It should be straightforward to extend to the biased
two-photon QRM, which is not shown here.

Trav\v{e}nec\cite{Trav} has used Braak's approach to solve this
model, but seems that the G-function like in Eq. (\ref{cerntral_2p})
is not given. Their coefficients are entangled in the two coupled
equations, which may prevent such a simple description for the
G-functions. In the present Eqs. ( \ref{coeef1}) and (\ref{coeef2}),
two coefficients are related one-by-one with the same index $n$,
which facilitates the derivations. This is also the advantage of
Bogoliubov operators, which result in free particle number
operators.

The  exceptional solutions to the two-photon QRM  have been studied
by Emary et al \cite{Bishop}. With the G- function in the two-photon
QRM at hand, we can also discuss the Juddian solution similar to the
one-photon model\cite{Braak}. The G function is also not analytic in
$x$ but has simple poles at $x=0,1,2....$. For special  values of
model parameters $g$, there are eigenvalues which do not correspond
to zeros of Eq. (\ref{cerntral_2p}); these are the exceptional
solutions. All exceptional eigenvalues are given by the positions of
the poles
\begin{equation}
E=\left( n+\frac 12\right) \sqrt{1-4g^2}-\frac 12
\end{equation}
which is exactly the same as that in Ref. \cite{Bishop}. The
necessary and sufficient condition for the occurrence of the
eigenvalue is $f_n(x=n)=0$, which provides a condition on the model
parameters $g$ and $\Delta $. They occur when the pole of $
G_{e,o}^{\pm }(x)$ at $x=n$ is lifted because its numerator in Eq.
(\ref{cerntral_2p}) vanishes. From Eq. (\ref{protion}), we know the
proportionality is justified only for the even or odd photonic
number respectively. The Juddian solutions are corresponding to
those states which are degenerate, and therefore are excluded within
this proportionality, so the lever crossing points with the same
even, or odd photonic numbers are corresponding to Juddian
solutions.

\section{Summary}

In this paper, by using the extended coherent states, Braak's exact solution
in the QRM is recovered explicitly in an alternative more physical way. A
preliminary extension to the two-photon QRM is also performed with the use
of extended squeeze states. The corresponding G-functions with the similar
form of Braak's G-function are derived, which has not been obtained before.
Both model can be treated in an unified way by the expansion in the Fock
space of the Bogoliubov operators. Further extensions to other more
complicated systems, such as multi-level, even multi-mode spin-boson model,
are straightforward, although perhaps a little bit tedious sometimes.

For multi-level spin-boson model, such as the finite-sized Dicke model\cite
{dicke}, the quantum chaos has been discussed\cite{Emary1}. We have expanded
the wavefunction in $N+1$ Bogoliubov operators for Dicke model with finite $%
N $ two-level atoms, and get numerically exact solutions previously\cite
{chenqh}. According to the above discussions and the link with Braak's
solutions, the exact solvability is ensured without doubt in this system.
The quasi-integrability and the quantum chaos in this system should be very
interesting. On the other hand, the multi-mode QRM has been also realized
experimentally in circuit QED systems\cite{Niemczyk}. The extensions to
these systems are in progressing.

\section{ACKNOWLEDGEMENTS}

We acknowledge useful discussions with Victor V. Albert, Daniel
Braak, I. Trav\v{e}nec, and Yu-Yu Zhang. This work was supported by
National Natural Science Foundation of China under Grant No.
11174254, National Basic Research Program of China (Grant Nos.
2011CBA00103 and 2009CB929104).

\end{document}